\documentclass[12pt,tightenlines,eqsecnum,floats,nofootinbib,amsmath,amssymb,aps,prd]{revtex4}
\usepackage{graphicx,verbatim}
\usepackage{amsmath}
\usepackage{amsfonts}
\usepackage{amssymb}
\def\Lie{\mathcal{L}}
\def\scri{\mathcal{I}}
\def\rmd{\mathrm{d}}
\def\L{\mathcal{L}}
\def\S{\mathcal{S}}
\def\T{\mathcal{T}}

\def\H{\mathcal{H}}

\def\Sbb{\mathbb{S}}
\def\Rbb{\mathbb{R}}

\newcommand{\pb}[1]{\hbox{\lower0.8ex\hbox{${}_{\leftarrow}$}}\kern-1.9ex{#1}}
\def\={\hat{=}}

\def\h{\hat}

\def\B{\mathfrak{B}}
\def\b{\mathfrak{b}}
\def\s{\mathfrak{s}}

\def\h{\mathfrak{h}}
\def\A{\mathfrak{A}}
\def\W{\mathfrak{W}}

\def\Gammabf{\mathbf{\Gamma}}
\def\Omegabf{\mathbf{\Omega}}

\def\be{\begin{equation}}
\def\ee{\end{equation}}
\def\ba{\begin{eqnarray}}
\def\ea{\end{eqnarray}}

\def\SO(3){\rm SO(3)}
\def\so(3){\rm so(3)}
\def\SO(4){\rm SO(4)}
\def\so(4){\rm so(4)}
\def\SO(1,4){\rm SO(1,4)}
\def\so(1,4){\rm so(1,4)}
\def\SU(2){\rm SU(2)}

\begin{document}

\title{Geometry and Physics of Null Infinity}
\author{Abhay Ashtekar}

\email{ashtekar@gravity.psu.edu} \affiliation{Institute for
Gravitation and the Cosmos \& Physics Department, Penn State,
University Park, PA 16802, U.S.A.}

\begin{abstract}

In asymptotically Minkowski space-times, one finds a surprisingly
rich interplay between geometry and physics in both the classical
and quantum regimes. On the mathematical side it involves null
geometry, infinite dimensional groups, symplectic geometry on the
space of gravitational connections and geometric quantization via
K\"ahler structures. On the physical side, null infinity provides a
natural home to study gravitational radiation and its structure
leads to several interesting effects such as an infinite dimensional
enlargement of the Poincar\'e group, geometrical expressions of
energy and momentum carried by gravitational waves, emergence of
non-trivial `vacuum configurations' and an unforeseen interplay 
between infrared properties of the quantum gravitational field and the 
enlargement of the asymptotic symmetry group. The goal of this article 
is to present a succinct summary of this subtle and beautiful interplay.
\end{abstract}

\maketitle

\section{Introduction}
\label{s1}

General relativity is widely regarded as the deepest realization of
the rich interplay between geometry and physics
known to date. The centerpiece of the theory is the fusion of the
gravitational field with space-time geometry. Consequently,
space-time geometry now becomes a dynamical entity and physics is
encoded in its properties. The most profound predictions of the
theory can be traced back to this interplay. It is because geometry
is unleashed from its fixed, rigid structure that the universe can
expand, we can have black holes and ripples of curvature can
propagate across cosmological distances carrying away energy and
momentum.

However, this very duality between geometry and gravity that
makes it difficult to develop mathematical tools that are necessary
to probe the general relativistic effects. An excellent example is
provided by isolated gravitating systems. In
the study of other fundamental interactions ---such as
electromagnetic--- we have a background Minkowski space-time at our
disposal, which is completely insensitive to the specific solutions
we study. Using this `inert', background geometry, then, it is easy
to introduce physically motivated boundary conditions, such as the
$1/r$ fall-off on fields, and arrive at well-defined expressions
of energy, momentum and angular momentum carried by electromagnetic
waves. In general relativity, the metric which gives meaning to the
`fall-off conditions' is itself the dynamical variable on which we
wish to impose the boundary conditions. This dual role creates
unfamiliar layers of complexity. It took some 4 decades after the
discovery of the theory to understand the boundary conditions at
spatial infinity and to define the total energy, momentum \cite{adm} 
and angular momentum \cite{aarh} of isolated systems \cite{spatial-review}. It took another 2 decades to develop the required geometric analysis to prove that, if the local energy density in matter is positive, then the total energy is also positive \cite{sy,ew}.

Gravitational radiation adds yet another level of subtleties. Already 
in 1917, Einstein had isolated the radiative modes of the gravitational 
field \emph{in the linear approximation} and derived the celebrated 
quadrupole formula. But subsequently, for several decades, there was a 
debate on whether this radiation was just an artifact of linearization. 
If one removes the anchor of a background Minkowski space-time 
geometry, can one still distinguish \emph{physical} gravitational waves 
from coordinate effects in \emph{full, non-linear} general relativity? 
It is interesting to note that at one stage Einstein himself thought
one could not. He wrote: ``Together with a young
collaborator I arrived at the interesting result that gravitational
waves do not exist, though they had been assumed to be a certainty
to the first approximation. This shows that non-linear gravitational
field equations tell us more or, rather, limit us more than we
had believed up to now" (see, e.g., \cite{er}).

It was only in the 1960s that Bondi, Sachs and their coworkers
resolved the long standing confusion by constructing a framework in
which the issue could be analyzed in an invariant fashion
\cite{bondi,sachs,null-review}. Specifically, because gravitational
waves travel within light cones in general relativity, they
constructed a systematic expansion of the metric as one moves away
from the sources in \emph{null} directions and studied asymptotics
at null infinity, in contradistinction with the Arnowitt Deser Misner 
(ADM) framework which focuses on spatial infinity. This construction 
was cast in a more convenient form through conformal techniques by 
Penrose \cite{rp} where null infinity is represented as the boundary, $
\scri$, of the physical space-time in its conformal completion. As this 
article will show, Penrose's $\scri$-framework has rich geometry containing important physics.

Starting from the late 1970s, the geometric analysis community has
studied the ADM framework and, more generally, the global elliptic
problems associated with the initial value formulation of Einstein's
equations in great detail (see, e.g., \cite{gms} for a review).
Similar investigations of the hyperbolic problems related to $\scri$
were undertaken later, using Einstein's equations on the physical 
metric \cite{dcsk}, as well as on the conformally rescaled metric which 
is well-behaved at $\scri$ \cite{hf1}. Over the years, there has been  
considerable progress \cite{lb,hf2}, and this area is likely to witness 
significant advances over the next decade (for the current status, see, e.g., \cite{sair}). 

On the physical side, the 
framework provides the conceptual foundation for a surprisingly large 
portion of gravitational science. In particular, it provides the basis 
for analytical approximation schemes as well as numerical simulations 
of the emission of gravitational waves in gravitational collapse and
binary coalescence. It lies at the heart of the definition of black
holes and is therefore important for all of the related mathematical
physics. (See, e.g., \cite{lb2,clp,bhs} for reviews.) Finally, in
quantum gravity, it provides a natural arena for the $S$-matrix theory
and is required in the formulation of conceptual issues such as that
of information loss during black hole evaporation and the role of
the CPT symmetry in quantum gravity (see, e.g., \cite{atv,apr,aa-asym,aa-bib}). On the physical side, interest in the structure of $\scri$ has increased  recently through more detailed analyses of, 
e.g., the `memory' effect \cite{memory1,memory2,memory3} in the 
classical theory, and a resurgence in the investigations of the 
representations of the Bondi Metzner Sachs (BMS) group \cite{bms-rep1,bms-rep2}, scattering amplitudes \cite{scattering1,scattering2} 
and the subtle infrared issues \cite{as,skinner,mcal} on the quantum 
side.

Therefore, an overview of geometry and physics of $\scri$ seems
appropriate at this juncture. Furthermore, structure of $\scri$
serves as a striking illustration of the power of the interplay
between these two disciplines that general relativity embodies, and
this is a jubilee volume celebrating the 100th anniversary of
Einstein's theory. However, because space is limited, this will be a 
broad overview without technical details or proofs. But a
special attempt has been made to address both mathematics and physics communities.

In section \ref{s2} we recall the basic structure of $\scri$,
including the enlargement of the Poincar\'e group to the infinite
dimensional BMS group. In section \ref{s3} we discuss the interplay
between geometry and physics that dominates the description of
radiative modes of the gravitational field in full general
relativity. Finally, in section \ref{s4} we discuss some of the
quantum issues. Our conventions are as follows. Space-time will be
assumed to be 4-dimensional and its metric will have signature
-,+,+,+. Physical fields will carry hats while those which are well
defined on the conformal completion will be unhatted. We will use
Penrose's abstract index notation \cite{rpwr} where the indices
serve as markers to denote the type of tensor field being considered
and do not refer to a chart.  The curvature tensors (in the
completion) are defined via: $2\nabla_{[a}\nabla_{b]} k_c =
R_{abc}{}^d k_d$, $R_{ac} = R_{abc}{}^b$ and $R = R_{ab}g^{ab}$. For
simplicity of presentation we will assume that all fields are
$C^\infty$. However, the main results discussed here hold for the weaker smoothness at $\scri$ assured by the Christodoulou-Klainnermann
analysis \cite{dcsk,lb}.

\section{The Bondi-Penrose framework}
\label{s2}

This section is divided into two parts. In the first we introduce
the relevant definitions of asymptotic flatness and summarize their
immediate consequences. In the second we discuss the structure of
the asymptotic symmetry group.

\subsection{Basic structure}
\label{s2.1}

Study  of null infinity is based on two basic definitions
\cite{rggh,aabs}. The first is weaker and suffices for the analysis
of gravitational radiation at $\scri$, while the second and stronger
becomes necessary to address global issues. In what follows, $\scri$
will stand \emph{either} for the future \emph{or} the past boundaries
$\scri^{\pm}$ of space-time. For the gravitational radiation theory
and black holes one is primarily interested in future null
infinity, $\scri^{+}$.\\

\textbf{Definition 1:} A space-time $(\hat{M},\hat{g}_{ab})$ will be
said to be \textit{asymptotically flat} at null infinity if there
exists a manifold $M$ with boundary $\scri$ equipped with a metric
$g_{ab}$ and a diffeomorphism from $\hat{M}$ onto $M\, \setminus\,
\scri$ (with which we identify $\hat{M}$ and $M\, \setminus\,
\scri$) such that:

i)\, there exists a smooth function $\Omega$ on $M$ with $g_{ab}=
\Omega^2 \hat{g}_{ab}$ on $\hat{M}$; $\Omega=0$ on $\scri$;\\
\indent\indent and $n_a := \nabla_a \Omega$ is nowhere vanishing on
$\scri$;

ii) $\scri$ is topologically $\Sbb^2\times \Rbb$; and,

iii)\,$\hat{g}_{ab}$ satisfies Einstein's equations $\hat{R}_{ab} -
\frac{1}{2} \hat{R} \hat{g}_{ab}  = 8 \pi G \; \hat{T}_{ab}$, where
$\Omega^{-2} \hat{T}_{ab}$ has \\ \indent\indent a smooth limit to $\scri$.\\

The first condition ensures that $(M,\,  g_{ab})$ is a conformal
completion of the physical space-time $(\hat{M},\, \hat{g}_{ab})$ in
which the boundary $\scri$ is at infinity with respect to the
physical metric $\hat{g}_{ab}$. The condition $\nabla_{a}\Omega
\not=0$ on $\scri$ ensures that $\Omega$ can be used as a coordinate
on $M$; we can perform Taylor expansions in $\Omega$ to capture the
degree of fall-off of physical fields. In terms of the physical
space-time $(\hat{M},\, \hat{g}_{ab})$, it ensures that $\Omega$
`falls-off as $1/r$',\, i.e., has the same asymptotic behavior as in
the standard conformal completion of Minkowski space-time. The
topological restriction captures the idea that one can move away
from the isolated system along null rays in any angular direction.
The last condition ensures that the matter fields fall-off
appropriately in the physical space-time $(\hat{M}, \hat{g}_{ab})$.
The specific fall-off of $\hat{T}_{ab}$ is motivated by
the analysis of test fields in Minkowski and Schwarzschild
space-times.

These conditions immediately imply that $\scri$ is necessarily a null, 
3-dimensional manifold. Therefore $n^a = \nabla^{a}\Omega$ is null and 
$\scri$ is ruled by the the integral curves of $n^a$, called the
\emph{generators}. The space $S$ of generators is topologically
$\Sbb^2$. The pull-back $q_{ab} := \pb{{g}_{ab}}$ of the conformally
rescaled metric to $\scri$ has signature 0,+,+ and is the lift to
$\scri$ of a positive definite metric $\bar{q}_{ab}$ on $S$.

Next, note that there is freedom to perform conformal rescalings: If
$\Omega$ is a permissible conformal factor for a physical space-time
$(\hat{M}, \hat{g}_{ab})$, so is $\Omega^\prime = \omega \Omega$
where $\omega$ is smooth on $M$ and nowhere vanishing on $\scri$.
Using this freedom, one can always choose a conformal completion
such that $\nabla_a n^a$ vanishes on $\scri$. We will always work
with such a \emph{divergence-free conformal frame}. The fall-off
condition iii) in Definition 1 implies that in these frames, a
stronger condition is automatically satisfied:
\be \label{divfree} \nabla_a n_b := \nabla_a\nabla_b \Omega\, \=\,
0. \ee
\emph{Here and throughout the rest of the paper,} whenever there may
be an ambiguity on whether a given equality holds on all of $M$ or just on $\scri$, \emph{we will use the symbol $\=$ to denote equality
restricted to} $\scri$. The remaining conformal freedom is given by
$\Omega^\prime = \omega\Omega$ where $\Lie_n \omega \= 0$. Eq.
(\ref{divfree}) implies that the torsion-free derivative operator
$\nabla$ compatible with $g_{ab}$ induces a torsion-free derivative
operator $D$ defined intrinsically on the 3-manifold $\scri$,
satisfying
\be \label{D} D_a q_{bc} = 0, \quad {\rm and}\quad D_a n^b = 0\, .
\ee
Finally, Definition 1 implies that the Weyl tensor $C_{abc}{}^d$ of
$g_{ab}$ vanishes on $\scri$. Therefore, the tensor field
\be \label{K} K_{abc}{}^d := \Omega^{-1}\, C_{abc}{}^d  \ee
has a smooth limit to $\scri$ and is called the \emph{leading order
Weyl tensor} at $\scri$.%
\footnote{Given a null frame \`a la Newman and Penrose
\cite{rp,rpwr}, the five complex functions $\Psi_4^o, \ldots
\Psi_0^o$ capture the 10 components of $K_{abc}{}^d$ at $\scri$. The 
`peeling' properties of the 5 NP scalars is a straightforward 
consequence of smoothness of $K_{abc}{}^d$ and the relation between the 
physical null tetrad defined by $\hat{g}_{ab}$ and the conformally 
rescaled one defined by $g_{ab}$. If differentiability at $\scri$ is weaker as in \cite{dcsk} and \cite{lb}, one has peeling only for some of the NP scalars.}

Fix any two divergence-free conformal frames $\Omega$ and
$\Omega^\prime$. Then, since the relative conformal factor satisfies
$\Lie_{n}\omega \=0$, the vector field $n^a$ is complete if and only
if ${n^\prime}^a$ is complete. Definition 1 does not demand
completeness of $\scri$ in the $\mathbb{R}$ direction. The second
definition does, i.e., it asks that $\scri$ have the same
\emph{global} structure as it has in the standard completion of
Minkowski space-time.\\

\textbf{Definition 2:} An asymptotically flat space-time is said to
be \emph{asymptotically Minkowski} if $\scri$ is complete in any
divergence-free conformal frame.\\

The second notion is important for black holes \cite{rggh}. Recall
that a space-time $(\hat{M}, \hat{g}_{ab})$ admits a black hole if
the past $J^-(\scri^+)$ of the future null infinity is a proper
subset of $\hat{M}$. Therefore, without the completeness
requirement, one could carry out a conformal completion of Minkowski
space-time which attaches to it just a `part of the $\scri^+$ of the
standard completion' and conclude that it admits a black hole
region! Completeness is also necessary in the discussion of the BMS
group (as opposed to the BMS Lie algebra). However, there are
examples (e.g. boost-symmetric space-times) in which this
requirement is not met but one can meaningfully discuss
gravitational radiation \cite{aabs}.

\subsection{The BMS Group}
\label{s2.2}

The asymptotic symmetry group $\B$ is the quotient ${\rm Diff}_{\infty}(M)/{\rm Diff}_{\infty}^{o}(M)$ of the group ${\rm
Diff}_{\infty}(M)$ of diffeomorphisms on the physical space-time
$(\hat{M}, \hat{g}_{ab})$ that preserve the boundary conditions
(imposed in Definition 2) by its subgroup ${\rm
Diff}_{\infty}^{o}(M)$ of diffeomorphisms that are asymptotically
identity. This is the BMS group. At the infinitesimal level,
elements of the Lie algebra $\b$ of $\B$ can be naturally
represented by vector fields $\xi^{a}$ on $\scri$,\, motions along
which preserve the \emph{universal structure,} i.e., the structure
that is shared by \emph{all} space-times satisfying Definition 2.

From our summary in the last sub-section, it follows that $\scri$ is
endowed with the following universal structure. It is topologically
$\Sbb^2 \times \Rbb$, and equipped with pairs of fields $(q_{ab},
n^a)$ such that: i) $q_{ab}$ is a degenerate metric of signature
0,+,+ with $q_{ab}n^{b} =0$ and $\mathcal{L}_{n}\,q_{ab} =0$; \, ii)
$n^{a}$  is complete;\, and, iii) any two pairs $(q_{ab}, n^a)$ and
$(q^\prime_{ab}, {n^\prime}^a)$ in the collection are related by a
conformal rescaling,
\be q^\prime_{ab} = \omega^2\,q_{ab} \quad {\rm and} \quad
{n^\prime}^a = \omega^{-1} n^a\, , \ee
where $\Lie_n\omega = 0$. Note that, because 2-spheres carry a
unique conformal structure, every $q_{ab}$  in this collection is
conformal to a unit 2-sphere metric.

It then follows that $\B$ can be characterized intrinsically in 
terms of $\scri$ as the subgroup of diffeomorphisms of $\scri$ that 
preserves this universal structure. It is simplest to explore its 
structure by working at the infinitesimal level. A vector field $\xi^a$ 
on $\scri$ will preserve the universal structure if and only if:
\be \Lie_{\xi} q_{ab} = 2\alpha\, q_{ab},\quad {\rm and} \quad
\Lie_{\xi} n^{a} = - \alpha\,Ä n^{a} \ee
for some function $\alpha$ on $\scri$ satisfying $\Lie_n\, \alpha
=0$. In particular, the vector fields $\xi^a = f n^a$ with
$\Lie_{n} f=0$ satisfy this condition. Furthermore, the subspace
$\s$ they form in the Lie algebra $\b$ of $\B$ is a Lie ideal in the
sense that $[\xi, \, f n^a] \in \s$ for all $\xi^a \in \b$ and all
$f n^a \in \s$. This is the Lie ideal of BMS
\emph{supertranslations.}

Next, note that the condition $\Lie_{\xi} n^{a} = \alpha n^{a}$ also
implies that \emph{every} BMS vector field $\xi^{a}$ can be
projected to a vector field $\bar{\xi}^{a}$ on the 2-sphere $S$ of
generators of $\scri$ which then characterizes the element of the
quotient $\b/ \s$ it naturally defines. Furthermore, the condition
that the space of pairs $(q_{ab}, n^{a})$ be preserved by the BMS
action implies that $\bar{\xi}^{a}$ is a conformal Killing field on
the space $S$ of generators of $\scri$, equipped with metrics
$\bar{q}_{ab}$. Since every $\bar{q}_{ab}$ is conformal to a round
2-sphere metric, it follows that the quotient  $\b/s$ is just the
Lie algebra of conformal isometries of a round 2-sphere. But this is
just the Lie algebra of the Lorentz group in 4 space-time
dimensions. Returning to finite diffeomorphisms, we conclude that
$\B$ is the semi-direct product, $\B = \S\, \ltimes \L$, of the
group $\S$ of supertranslations with the Lorentz group $\L$. Thus,
although $\B$ is infinite dimensional, its structure mimics that of
the Poincar\'e group, the 4-dimensional Abelian group of
translations being replaced by that by the infinite dimensional
Abelian group $\S$ of supertranslations. Because each
supertranslation is represented by a vector field of the type $f
n^a$ on $\scri$ where $f$ is the pull-back to $\scri$ of a function
$\bar{f}$ on the 2-sphere $S$ of generators, supertranslations can be 
regarded as `angle dependent translations' in the physical space-time. 
While the Poincar\'e group admits a 4-parameter family of Lorentz 
groups, each labeled by a point in Minkowski space-time, the BMS group 
has an infinite number of Lorentz subgroups, each labeled by a 2-sphere 
cross section of $\scri$ (i.e., a lift of $S$ to $\scri$).

However, because the space of generators of $\scri$ is topologically
$\Sbb^2$, one can show that $\B$ also admits a unique normal,
Abelian subgroup $\T$ \cite{sachs}. This is a subgroup of $\S$ and
in Minkowski space-time it coincides with the group of space-time
translations. Because there is a canonical translation subgroup
$\T$, the notion of 4-momentum continues to be well-defined for
gravitational waves, and also for matter waves in the presence of
gravitational radiation. But the familiar notion of angular-momentum
we use in Minkowski space-time now acquires an infinite dimensional,
supertranslation ambiguity. Thus, the enlargement from the
Poincar\'e to the infinite dimensional BMS group has direct physical
consequences but they are subtle. We will see in section \ref{s3}
that the enlargement is directly related to the presence of
gravitational waves.

Finally, as one would expect, every Killing vector $K^a$ of the
physical space-time satisfying Definition 1 extends naturally to a
BMS vector field on $\scri$. Furthermore, if the leading order Weyl
curvature is such that $K_{abcd}n^d$ is not identically zero at
$\scri$, then the isometry group is at most 4-dimensional, and a
subgroup of the isometry group of the Schwarzschild space-time. If
the extension of $K^a$ is a supertranslation on $\scri$, then it is
necessarily a \emph{time translation} in $\T$ \cite{aabs}.\\

\emph{Remark:} It is sometimes convenient to further restrict the
conformal freedom at $\scri$ by demanding that the metric
$\bar{q}_{ab}$ on the space $S$ of generators be a unit 2-sphere
metric. This is always possible and these conformal frames are
called \emph{Bondi frames}. In a Bondi frame, a BMS supertranslation
$\xi^a = fn^a$ is a translation if and only if
\be\label{trans} D_aD_b f  \propto\, q_{ab}\, .  \ee
This is equivalent to asking that $f$ be a linear combination of the
first four spherical harmonics, $f = f_0\,Y_{0,0}(\theta,\phi) +
f_m\, Y_{1,m}(\theta,\phi)$ for some real constants $f_0, f_m\,\,
({\rm with}\,\, m =-1,0,1) $. However, restricting oneself to Bondi
frames can also make certain conceptual issues more difficult to
analyze, especially in the definition of the `BMS charges'
---which generalize the notions of energy-momentum and angular
momentum in view of the enlargement of the Poincar\'e group to the
BMS--- and the fluxes of BMS momenta across $\scri$ \cite{td}.

\section{Radiative modes in exact general relativity}
\label{s3}

This section is divided into three parts. In the first, we explore
higher (asymptotic) order geometrical structures that are not
universal and encode physical information of the given space-time;
in the second,  we introduce the notion of `vacuum configurations'
following gauge theories, and in the third we discuss the symplectic
geometry of radiative modes and momentum maps defined by the BMS
symmetries.

\subsection{Further geometrical structures at $\scri$}
\label{s3.1}

The universal structure of $\scri$ is common to \emph{all}
asymptotically Minkowski space-times. Therefore we will regard the
$\Sbb^2\times \Rbb$ topology and the collection of pairs $(q_{ab},
n^a)$ as the \emph{zeroth order} structure of $\scri$. The
\emph{first order} structure is the connection $D$ defined
\emph{intrinsically} on $\scri$ in any given conformal completion, induced by the torsion-free connection $\nabla$ compatible with 
$g_{ab}$. As we will see, it encodes the `radiative information' in the
physical space-time $(\hat{M}, \hat{g}_{ab})$ and therefore varies
from one space-time to another. In particular, Eqs. (\ref{D}) do not
determine $D$ uniquely because $q_{ab}$ is degenerate. 

It is convenient to first fix a conformal frame $(q_{ab}, n^a)$ on
$\scri$, develop the framework, and then study what happens under
conformal rescalings $g_{ab} \to g^\prime_{ab} = \omega^2 g_{ab}$.
However, even when $\omega = 1$ on $\scri$ so that $q_{ab}$ and
$n^a$ are unchanged, $D$ transforms via $D^\prime_a k_b = D_a k_b +
f\, (n^m k_m)\, q_{ab}$, where $k_b$ is an arbitrary 1-form on
$\scri$ and $f$ is defined via $ \nabla^a \omega \= f n^a$.
Therefore we are led to introduce an equivalence relation:
\be \label{equiv} D \approx D' \quad {\rm iff}\quad  (D^\prime_a -
D_a) k_b = (f\, n^ck_c)\, q_{ab} \ee
where $f$ is an arbitrary function on $\scri$. We will denote the
equivalence class by\, $\{D\}$. Eqs. (\ref{D}) now imply that any
two connections in a given conformal frame $(q_{ab}, n^a)$ are
related by $(D^\prime_a - D_a) k_b = \Sigma_{ab}\, n^c k_c$ where
$\Sigma_{[ab]}= 0$ and $\Sigma_{ab} n^b =0$. Therefore the
difference $\{D^\prime\}- \{D\}$ between the equivalence classes of
connections is completely characterized by the trace-free part
\be \sigma_{ab} = \Sigma_{ab} - \frac{1}{2}\, \Sigma_{cd}q^{cd}\,
q_{ab}\ee
of $\Sigma_{ab}$ (where $\Sigma_{cd}q^{cd}$ is well defined because
$\Sigma_{ab} n^b =0$). The space $\Gammabf$ of equivalence classes
$\{D\}$ is an affine space. Hence by choosing any $\{D\}_{0}$ as
origin, any $\{D\}$ can be uniquely labeled by a symmetric,
trace-free tensor field $\sigma_{ab}$ on $\scri$ which is
\emph{transverse}, i.e., satisfies $\sigma_{ab} n^b =0$. Physically,
the two independent components of $\sigma_{ab}$ can be regarded as
representing the two radiative degrees of freedom of the
gravitational field in exact general relativity \cite{aa-rad}. It is
striking that, in spite of the complicated non-linearities of
general relativity, the structure of the radiative modes is so
simple at $\scri$. In a characteristic initial value problem, $\{D\}$
represents the part of the initial data that is freely specifiable
on 3-dimensional $\scri$ (in addition to other quantities that have
to be specified on a transverse null surface $\mathcal{N}$ and the 2-
sphere at which $\mathcal{N}$ intersects $\scri$) 
\cite{characteristic}.

Since $\scri$ is 3-dimensional, the curvature tensor $R_{abc}{}^d$
of any $D$ is completely determined by a second rank tensor
$S_a{}^b$:
\be R_{abc}{}^d\, = \, q_{c[a} S_{b]}{}^d + S_{c[a} \delta_{b]}{}^d
\, ,\ee
where $S_{ab} = S_a{}^c q_{bc}$. Because of Eq. (\ref{D}), the field
$S_{a}{}^b$ on $\scri$ satisfies $S_{[ab]} =0,\, S_{ab}q^{ab} =
\bar{R}$ and $S_{a}{}^bn^a \propto n^b$, where $\bar{R}$ is the
pull-back to $\scri$ of the scalar curvature of $\bar{q}_{ab}$ on
the 2-sphere of generators $S$. One would expect $R_{abc}{}^{d}$ to
carry physically important information about gravitational waves in
the given space-time. This is indeed the case modulo a small
complication. Let us first consider $S_{ab}$. The complication is
that if we change the conformal frame, the curvature tensor
transforms in a complicated manner. Thus, part of this curvature is
`gauge' in the sense that it does not carry any physical
information. If $q^\prime_{ab} = \omega^2 q_{ab},\, {n^\prime}^a =
\omega^{-1}\, n^a$ (with $\omega$ not necessarily equal to 1 at
$\scri$) we have:
\begin{align}  {D^\prime}_ak_b\, & \=\, D_ak_b - 2 \omega^{-1} k_{(a}D_
{b)}\omega + \omega^{-1} (\nabla^{m}\omega)\, k_m\, q_{ab}\\
S^\prime_{ab}\, & \= \,S_{ab} - 2\omega^{-1} D_a D_b \omega + 4 \omega^
{-2}D_a\omega D_b\omega - \omega^{-2}\, (q^{mn}D_m\omega\, D_n\omega)\, 
q_{ab}\, .
\end{align}
($q^{mn}D_m\omega\, D_n\omega$ is well-defined because $\Lie_n
\omega =0$). The task is to extract conformally invariant
information from this rather complicated transformation property.
This can be done in an elegant manner thanks to a result due to
Geroch \cite{rg}: In any conformal frame $(q_{ab}, n^a)$, there is a
unique `kinematical' tensor field $\rho_{ab}$ on $\scri$ satisfying:
\be \rho_{[ab]} = 0; \,\,\,\, \rho_{ab}n^b =0; \,\,\,\,
\rho_{ab}q^{ab} =\bar{R}\,\,\,\, {\rm and}\,\,\,\, D_{[a}\rho_{b]c}
=0\, .\ee
One can show that under conformal rescalings, the transformation
property of $\rho_{ab}$ mimics that of $S_{ab}$ so that
\be N_{ab} := S_{ab}\, -\,  \rho_{ab} \ee
\emph{is conformally invariant}. Thus, the role of $\rho_{ab}$ is
simply to remove from $S_{ab}$ the gauge-dependent part. $N_{ab}$ is
called the \emph{Bondi news tensor} and represents the `second
order' structure at $\scri$, constructed from the derivative of the
first order structure encoded in $\{D\}$.

However, it is $S_a{}^b$ rather than $S_{ab}$ that has the full
information of the curvature of $\{D\}$. (One cannot reconstruct
$S_a{}^b$ from $S_{ab}$ because $q_{ab}$ is degenerate.) Can we also
extract the gauge-independent part from full $S_a{}^b$? This is
indeed possible using the field $K^{abcd}$ of Eq. (\ref{K}) that
captures the leading order, asymptotic space-time curvature. (Here
the indices are of course raised using $g^{ab}$). The `magnetic
part'
\be {}^\star \!K^{ac}\, :\= \,{}^\star\!K^{abcd} n_b n_d \ee
of this asymptotic curvature is tangential to $\scri$. It turns out
that this part is related to $S_a{}^b$ via
\be {}^\star\!K^{ac} = 2\epsilon^{mna}D_m S_n{}^{c}\, ,  \ee
where $\epsilon^{abc} = \epsilon^{abcd}n_d$ is the intrinsic volume
3-form on $\scri$. Since ${}^\star\!K^{ac}$ involves derivatives of
$S_a{}^b$, it represents the third order structure at $\scri$. By
definition it is also symmetric and trace-free.%
\footnote{In a Newman-Penrose null tetrad \cite{rp,rpwr}, components
of ${}^\star\!K^{ac}$ are given by $\Psi_4^o, \Psi_3^o, {\rm
Im}\Psi_2^o$. Note that the components do not include ${\rm
Re}\Psi_2^o$, which represents the `longitudinal mode' of the
gravitational field and determines the mass in stationary
space-times. For detailed derivations of the results quoted in this
sub-section, see \cite{aa-rad,aa-asym}.}
Under conformal rescalings, ${}^\star\!K^{ab}$ transforms
covariantly:
\be \label{Ktrans} {}^\star\!{K^\prime}^{ab} = \omega^{-5}\,
{}^\star\!K^{ab}\, \ee
Finally one can show locally, i.e., without having to refer to
completeness of $\scri$, that if ${}^\star\!K^{ab} =0$ then $N_{ab}
=0$ \cite{aabbak}. ${}^\star\!K^{ab}$ captures the full gauge
covariant information in the radiative mode $\{D\}$, in the same
sense that the (Lie algebra-valued) curvature 2-form $F_{ab}$
captures the full gauge covariant local information in a non-Abelian
connection 1-form $A_a$.

\subsection{Vacuum configurations and the Poincar\'e reduction}
\label{s3.2}

If ${}^\star\!{K}^{ab} =0$, the equivalence class $\{D\}$ of
connections is said to be \emph{trivial}. It is clear from Eq.
(\ref{Ktrans}) that this is a conformally invariant notion. In what
follows, for brevity, we will drop the words `equivalence
class' and refer to $\{D\}$ simply as a connection. Following the
terminology used in gauge theories in which the dynamical variable
is again a connection, trivial connections are referred to as
\emph{`vacuum configurations'}, and denoted by $\{\mathring{D}\}$.
If a physical space-time $(\hat{M}, \hat{g}_{ab})$ is such that the
induced connection on $\scri$ is trivial, then it does not contain
gravitational waves. In particular, as we will see below, there is
no flux of energy or momentum or angular momentum at $\scri$. All
asymptotically flat, stationary space-times induce a trivial
connection on $\scri$. More generally, in physics one is interested
only in those space-times in which gravitational radiation falls-off
sufficiently fast in the distant past and in the distant future.
Thus, all configurations $\{D\}$ of physical interest tend to vacuum
configurations $\{\mathring{D}^{\mp}\}$ as we move to spatial
infinity $i^o$ and future time-like infinity $i^+$ along $\scri^+$.

Let $\mathring{\Gammabf}$ denote the subspace of $\Gammabf$ spanned
by vacuum configurations.  We will now explore the structure of
$\mathring{{\mathbf{\Gamma}}}$ and the action of the BMS group $\B$
on it following \cite{aa-rad}. As we noted in section \ref{s3}, the difference $\{{D}\}^\prime -
\{D\}$ between any two connections in $\Gammabf$ is characterized by
a trace-free symmetric tensor field $\sigma_{ab}$ that it transverse
to $n^a$. One can show that if $\{\mathring{D}\}^\prime$ and
$\{\mathring{D}\}$ are both in $\mathring{\Gammabf}$  the difference
has the following form in any Bondi conformal frame:
\be \label{sigma1}\sigma_{ab} = D_a D_b s - \frac{1}{2}\, (q^{cd}D_c
D_d s)\, q_{ab} \, \quad \hbox{\rm for some function $s$ on $\scri$ with} \quad \Lie_n s =0\, . \ee
Since we are in a Bondi frame where $\bar{q}_{ab}$ is the unit
2-sphere metric, the right side vanishes if and only if $s$ is the
lift to $\scri$ of $\bar{s}$ on $S$ of the form $\bar{s} = s_0
Y_{0,0} + s_m Y_{1,m},\,\, m = -1,0,1$, where $s_0$ and $s_m$ are constants. Thus,
there are `as many' vacuum configurations as there are elements of
$\mathcal{F}/\mathcal{F}_{0,1}$ of the quotient of the space
$\mathcal{F}$ of functions on $S$ by the 4-dimensional space $\mathcal
{F}_{0,1}$ spanned by the first 4 spherical harmonics. Next, recall 
that a supertranslation is represented by a BMS vector field $\xi^a$ of 
the form $\xi^a = fn^a$ with $\Lie_n f =0$. Under the 1-parameter 
family of diffeomorphisms $d(\lambda)$ it generates, a connection $\{D
\}$ is mapped to $\{D\}(\lambda)$. The difference $\{D\}(\lambda) -
\{D\}$ is characterized by a 1-parameter family of tensor fields
$\sigma_{ab}(\lambda)$ which, in a Bondi frame, has the form
\be \label{sigma2} \sigma_{ab}(\lambda) = \lambda\, \big(D_a D_b f -
\frac{1}{2} (q^{cd}D_cD_d f)\, q_{ab}\big) \ee
Eqs. (\ref{sigma1}) and (\ref{sigma2}) imply that:\\
(i) The action of supertranslations preserves the subspace
$\mathring{\Gammabf}$ of $\Gammabf$ and the action is transitive;\\
(ii) A vacuum configuration $\{\mathring{D}\}$ is left invariant
under this action if and only if $\xi^a$ is a BMS \emph{translation}
and BMS translations leave \emph{each} vacuum configuration
invariant.\\
Thus, the quotient $\mathcal{S}/\mathcal{T}$ of the group
$\mathcal{S}$ of BMS supertranslations by its 4-dimensional normal
subgroup $\mathcal{T}$ of translations acts simply and transitively
on the space $\mathring{\Gammabf}$.  Note however that while this
implies that $\mathcal{S}/\mathcal{T}$ is isomorphic to
$\mathring{\Gammabf}$, there is no \emph{natural} isomorphism
between them.

The discussion of section \ref{s2.2} implies that the BMS group $\B$
admits `as many' Poincar\'e subgroups $\mathcal{P}$ as there are
elements of  $\mathcal{S}/\mathcal{T}$. More precisely, the group
$\mathcal{S}/\mathcal{T}$ acts simply and transitively on the space
${\mathbf{\Pi}}$ of Poincar\'e subgroups $\mathcal{P}$ of $\B$.
Therefore it is natural to ask if there is a natural isomorphism
between the two representation spaces, $\mathring{\Gammabf}$ and
${\mathbf{\Pi}}$, on both of which $\mathcal{S}/\mathcal{T}$ acts
simply and transitively. The answer turns out to be in the
affirmative.

Let us fix an element $\{\mathring{D}\}$ of $\mathring{\Gammabf}$
and ask for the subgroup of the BMS group that leaves it invariant.
A detailed analysis shows that it is precisely a Poincar\'e subgroup
$\mathcal{P}_{\{\mathring{D}\}}$ of $\B$ \cite{aa-rad} (see also
\cite{etnrp}). Furthermore, this map from $\mathring{\Gammabf}$ to
${\mathbf{\Pi}}$ is an isomorphism. Thus, \emph{the enlargement of
the Poincar\'e to the BMS group at null infinity is directly related
to the `vacuum degeneracy'}, i.e., to the fact that there are
distinct connections $\{\mathring{D}\}$ which have trivial curvature
in the sense that ${}^\star\!K^{ab}$ they define vanishes
identically (and therefore their Bondi news also vanishes). There is
a precise sense in which this structure is analogous to that
encountered in non-Abelian gauge theories in Minkowski
space-time. However, there are also some key differences.
Similarities and differences are both spelled out in \cite{aa-rad}.

From a gravitational physics perspective, the situation can be
summarized as follows. As we already noted above, physically, one is
interested in space-times which induce connections $\{D\}$ on
$\scri$ which asymptotically tend to vacuum configurations
$\{\mathring{D}^{\mp}\}$ as we move along $\scri^{+}$ to the past
towards $i^o$ and to the future towards $i^+$. For physically
reasonable sources, generically $\{\mathring{D}^\mp\}$ are
\emph{distinct} vacuum configurations. The difference encodes both the 
`linear and non-linear'  or  `ordinary and null' memory effects \cite
{memory2,memory3}. If we could restrict ourselves to space-times for 
which $\{\mathring{D}^{-}\}= \{\mathring{D}^{+}\}$, then we could add 
this vacuum configuration to the universal structure at $\scri$ and we
would then have a reduction of the BMS group to the Poincar\'e
group. But this would be too severe a restriction and we would be
left with a very special class of isolated gravitating systems.
Thus, if we wish to construct a framework that incorporates a
sufficiently large class of physically interesting sources emitting
gravitational waves, we have to live with the infinite `vacuum
degeneracy' and the corresponding enlargement of the Poincar\'e
group to the infinite dimensional BMS group.\\

\emph{Remark:} Our entire discussion involves $\scri$ and is
therefore formulated in terms of a conformal completion of the
physical space-time. In terms of the physical space-time itself, the
main point can be summarized as follows. Suppose we were to consider
a class of space-times $(\hat{M}, \hat{g}_{ab})$ which are such that
we can extract a canonical Minkowski metric $\hat{\eta}_{ab}$
asymptotically and write $\hat{g}_{ab} = \hat{\eta}_{ab} + (1/r)\,
\hat{h}_{ab}$. Then we could use the Poincar\'e group selected by
$\hat{\eta}_{ab}$ as our asymptotic symmetry group. However, if we
allow generic sources of gravitational waves, then the presence of
radiation in the asymptotic region prevents us from choosing a
canonical $\hat{\eta}_{ab}$. Given one such $\hat{\eta}_{ab}$, we
can perform a diffeomorphism corresponding to an angle-dependent
translation and produce another $\hat{\eta}^\prime_{ab}$ which is
equally good for our expansion. The Poincar\'e groups $\mathcal{P}$
and $\mathcal{P}^\prime$ selected by the two Minkowski metrics would
then fail to agree even to leading order. Their
translation subgroups \emph{would} agree asymptotically, but the
two sets of Lorentz groups will differ, and the difference will
carry the imprint of the supertranslation relating $\hat
{\eta}_{ab}$ and $\hat\eta^{\prime}_{ab}$. Returning to
the language of null infinity, each Minkowski metric will enable us
to select a 4-parameter family of cross-sections of $\scri$ related
by BMS translations. That is why the translation groups agree.
However, the unprimed and the primed families of cross sections
would be distinct, and the two families would be related by a
supertranslation. Each family would determine a trivial connection
in $\mathring{\Gammabf}$ canonically, but the two `vacua'
$\{\mathring{D}\}$ and $\{\mathring{D}\}^\prime$ would be distinct
and select distinct Poincar\'e subgroups $\mathcal{P}$ and
$\mathcal{P}^\prime$ of the BMS group.

\subsection{Symplectic geometry of radiative modes}
\label{s3.3}

So far, we have regarded $\scri$ as the conformal boundary in the
completed space-time $(M, g_{ab})$ and used fields in $M$ to induce
fields on $\scri$. Now we wish to construct the phase space of all
possible radiative modes of the gravitational field in full general
relativity. Therefore we will now regard $\scri$ as an abstract
manifold, topologically $\mathbb{S}^2\times \mathbb{R}$, not
attached to any physical space-time. It will be endowed with: i) the
pairs $(q_{ab}, n^a)$, where $n^a$ is complete as in section
\ref{s2.1} (i.e., the universal structure); and, ii) connections $D$
that are compatible with the pairs $(q_{ab}, n^a)$ in the sense of
section \ref{s3.1}.

It is convenient to first fix a conformal frame $(q_{ab}, n^a)$ and,
without any reference to a space-time, just compute the curvature
$R_{abc}{}^d$ of any given $D$, extract $S_a{}^b$, introduce the
kinematical tensor $\rho_{ab}$, and define the news tensor $N_{ab}$
and the field ${}^\star\!K^{ab}$ using equations displayed in
section \ref{s3.1}. These fields continue to satisfy all the
relations at $\scri$ we discussed even though $\scri$ is not the
boundary of any space-time. We can then introduce the equivalence
relation (\ref{equiv}) among connections $D$ and show that $N_{ab}$
and ${}^\star\!K^{ab}$ depend only on the equivalence class.
Finally, we can relate the structures associated with any two
conformal frames $(q_{ab}, n^{a})$ and $(q^\prime_{ab} = \omega^2
q_{ab}, \, {n^\prime}^a = \omega^{-1} n^a)$, by specifying the
transformation property on the equivalence classes $\{D\}$:
\be (\{D_a\}^\prime - \{D_a\})k_b \approx 2 \omega^{-1} k_{(b}\,
D_{a)}\omega\, . \ee
where $\approx$ stands for equality modulo additive terms of the
type $f q_{ab}$ for some $f$. We can then establish the
transformation properties $N^\prime_{ab} = N_{ab}$ and
${}^\star\!{K^\prime}^{ab} = \omega^{-5} {}^\star\!K^{ab}$. Thus,
all the structure that refers to the `radiative modes' can be 
introduced directly on an abstractly defined $\scri$ without reference 
to a physical space-time \cite{aa-rad,aams,aa-asym}.

The phase space $\Gammabf$ will consist of the connections $\{D\}$
subject to the condition that they approach \emph{some} vacuum
configurations $\{\mathring{D}^\mp\}$ at an `appropriate
rate' as one approaches the two `ends', $i^o$ and $i^+$ of $\scri$.%
\footnote{To make $\Gammabf$ into a proper symplectic space we need
to specify function spaces more precisely. But that would require a
long detour, and is beyond the scope of this brief report. One
avenue, which involves $C^\infty$ fields (defined intrinsically on
$\scri$), is discussed in \cite{aams}. But there should be simpler
options if one considers only $C^k$ fields in the spirit of the global
existence results of the Christodoulou-Klainnerman type. This is an interesting open issue.}
This is the space of radiative modes at $\scri$. To define the
symplectic structure $\Omegabf$ on $\Gammabf$, we first fix a
conformal frame $(q_{ab}, n^a)$ on $\scri$.  Then, at any point
$\{D\}$ of $\Gammabf$, the symplectic structure is a
(weakly-non-degenerate) 2-form  that assigns to any two tangent
vectors $\sigma_{ab}, \sigma^\prime_{ab}$ at $\{D\}$ the following
number:
\be \label{symp} \Omegabf\mid_{\{D\}}\,(\sigma, \sigma^\prime) \, :=
\frac{1}{8\pi G}\, \int_{\scri} \big[\sigma_{ab}\, \Lie_{n}
\sigma^\prime_{cd}\, -\, \sigma^\prime_{ab}\, \Lie_{n} \sigma_{cd}
\big]\, q^{ac}\, q^{bd}\, \epsilon_{mnp}\, {\rmd}S^{mnp}\, . \ee
Here $G$ is Newton's constant and $\epsilon_{abc}$ is the 3-form on
$\scri$ (unique up to sign) defined as follows: it is the inverse of
the antisymmetric tensor field on $\scri$ satisfying
\be \epsilon^{mnp}\, \epsilon^{abc} q_{ma}\, q_{nb} = n^p n^c \, \ee
(so that $\epsilon^{mnp}\, \epsilon_{mnp}\, = \, 6$). Since
$n^a\sigma_{ab} =0$, it follows that the right side of (\ref{symp})
is independent of the particular choice of the `inverse' $q^{ab}$ of
$q_{ab}$ made in this evaluation. Finally, under conformal
rescalings $(q^\prime_{ab} = \omega^2 q_{ab}, \, {n^\prime}^a =
\omega^{-1} n^a)$, one finds $\sigma^\prime_{ab} = \omega
\sigma_{ab},\, {q^\prime}^{ab} = \omega^{-2}\, q^{ab}$ and
$\epsilon^\prime_{abc} = \omega^3\, \epsilon_{abc}$, so that the
right hand side of (\ref{symp}) remains unchanged. Hence the
symplectic structure is insensitive to the initial choice of the
conformal frame.

One can show that under the action of the BMS group, the symplectic
structure is left invariant and one can therefore calculate the
Hamiltonians generating these symplectomorphisms. This is a rather
subtle and technically difficult calculation. However, the final
result is rather simple \cite{aams}: the Hamiltonian corresponding
to the BMS vector field $\xi^a$ is given by
\be \label{H} H_{\xi} (\{D\})\, = \, \frac{1}{16\pi G}\,
\int_{\scri}\, \big[ N_{ab}\, (\Lie_{\xi}\, D_c - D_c \Lie_{\xi})\,
\ell_d\,  +\, 2N_{ab}\, \ell_c D_d \alpha \big]\, q^{ac} q^{bd}\,
\epsilon_{mnp}\, {\rmd}S^{mnp}\, , \ee
where $\ell_a$ is any 1-form on $\scri$ satisfying $\ell_an^a = -1$
and the function $\alpha$ is define via $\Lie_\xi q_{ab} = 2\alpha\,
q_{ab}$. The right side is independent of the particular choice of
$\ell_a$, of the `inverse' $q^{ab}$ of $q_{ab}$ and of the conformal
frame $(q_{ab}, n^a)$ made in its evaluation.  Mathematically,
(\ref{H}) provides a mapping from $\Gammabf$ to the dual of the BMS
Lie algebra $\b$; i.e., it is a momentum map associated with the
natural action of the BMS group on the phase space $(\Gammabf,
\Omegabf)$. Physically, $H_\xi$ represents the flux of the component
of the `BMS momentum' associated with the BMS generator $\xi^a$
across $\scri$.

This momentum map has several interesting features. The salient among 
them can be summarized as follows.

i) Since the news tensor $N_{ab}$ vanishes at any vacuum
configuration $\{\mathring{D}\}$, the flux of the entire BMS
momentum across $\scri$ vanishes identically in this case. This is just 
what one would expect of a `classical vacuum'.

ii) If $\xi^a$ is a supertranslation, $\xi^a= fn^a$, then $\alpha=0$
and so the second term in the integrand vanishes and the first term
simplifies to
\be H_{\xi}(\{D\}) \, =\, \frac{1}{16\pi G}\,\int_{\scri} N_{ab}[ f
S_{cd} + D_c D_d f] q^{ac}\, q^{bd}\, \epsilon_{mnp}\,
{\rmd}S^{mnp}\, . \ee
In a Bondi frame, $S_{cd} = N_{cd} + \textstyle{\frac{1}{2}}\,
q_{cd}$ and hence
\be \label{Hst}H_{\xi}(\{D\}) \, =\, \frac{1}{16\pi G}\,\int_{\scri}
N_{ab}[ f N_{cd} + f q_{cd} + D_c D_d f ]\, q^{ac}\, q^{bd}\,
\epsilon_{mnp}\,  {\rmd}S^{mnp}\, . \ee

iii) Recall that $D_c D_d f +  f q_{cd}$ vanishes in a Bondi frame if 
and only if $fn^a$ is a BMS translation. So, in that case, the 
Hamiltonian simplifies further to
\be \label{Ht} H_{\xi}(\{D\}) \, =\, \frac{1}{16\pi G}\,\int_{\scri}
\,[f N_{ab} N_{cd}]\, q^{ac}\, q^{bd}\, \epsilon_{mnp}\,
{\rmd}S^{mnp}\, . \ee
This is the celebrated expression of the flux of energy-momentum at
null infinity that Bondi and Sachs had proposed in the 1960s. For a
(future-directed) time translation, $f$ is positive and so the flux
of energy is manifestly positive. They extracted this expression
using the anticipated asymptotic behavior of the metric as one moves
away from the sources in null directions, and the asymptotic field
equations governing metric coefficients. It is this proposal that
served to resolve the then controversial issue of whether gravitational 
waves are physical. As Bondi is said to have put it, they are real 
because they carry energy and we can heat water with them! In our 
discussion, the expression arises from principles of symplectic 
geometry, i.e., as the Hamiltonian generating BMS translations. This is 
a striking example of the rich interplay between geometry and physics 
that one encounters repeatedly in general relativity.

We will conclude with a discussion of an interesting subtlety that 
further illuminates the interplay between geometry and physics. In the 
early literature, it was thought that the flux of supermomentum is also
given by Eq.(\ref{Ht}), where now $f$ is a general function on the
2-sphere $S$ of generators of $\scri$, rather than a BMS translation. 
But a careful analysis via symplectic geometry revealed that, for a 
general supertranslation, there is an extra term $ D_c D_d f + 2 f q_
{cd}$, as in the integrand on the right side Eq. (\ref{Hst}). In the
older work, the phase space was constructed from `shear tensors'
rather than from connections and was thus regarded as a vector space
rather than an affine space. This effectively amounts to fixing a
$\{\mathring{D}\}$ once and for all as the origin and replacing
every other $\{D\}$ with the tensor $\sigma_{ab}$ relating that
$\{D\}$ to the fiducial $\{\mathring{D}\}$. However, as we discussed
in section \ref{s3.3}, supertranslations act non-trivially on vacuum
configurations and therefore the fiducial $\{\mathring{D}\}$ is
mapped to another $\{\mathring{D}^\prime\}$ under the action of
$\mathcal{S}/\mathcal{T}$, invalidating the use of a fixed vector
space structure in the older treatment. There are checks in terms of
`balance laws' that establish that the extra term arising from the
shift of the origin is essential \cite{aams,td}. Thus, the fact that the phase
space has the structure of an affine space rather than a vector
space has direct physical consequences.

Finally the phase space $(\Gammabf, \Omegabf)$ and the subsequent
constructions of the Hamiltonians are completely parallel to those
used in the analysis of Maxwell and Yang-Mills fields in Minkowski
space-time at $\scri$; the only difference is that the symmetry
group is simply the Poincar\'e group rather than the BMS. In 
these examples, if there are no sources (i.e., if we have `pure'
gauge fields), one can show that the phase space of radiative modes
is isomorphic with the `covariant phase space' constructed from the
space of solutions \cite{aams}. The same result has been obtained
for source-free solutions to Einstein's equations \cite{aaam} but
without a rigorous treatment of function spaces. In light of
results on global existence for `small' data \cite{dcsk,lb} that have
been established since then, it should be possible to put results of
\cite{aaam} on a sounder mathematical footing. This extension would
be of considerable interest both for classical and quantum gravity.

\section{Quantum aspects}
\label{s4}

Null infinity is especially well-suited to discuss the $S$-matrix 
theory of zero rest mass fields. In the
case of the Maxwell field in Minkowski space-time, one can construct
the Fock space of quantum states directly from the phase space of
radiative modes at $\scri^\mp$, without reference to the interior of
space-time. They serve as the `in' and `out' photon states for the
$S$-matrix. Furthermore, the description at $\scri$ provides a
clean, geometric understanding of the subtle infra-red problems
\cite{fk} one encounters in the scattering theory in quantum
electrodynamics \cite{aa-asym}.

In the gravitational case, the situation is more complicated, first
because of the absence of a Minkowski space-time in the background,
and second because the gravitational
field acts as its own source. The $\scri$-framework enables one to
surmount the first difficulty in that one can construct the
asymptotic Hilbert spaces of states on $\scri^\mp$ for \emph{full
non-linear general relativity}. Currently significant advances are
being made in calculating the $S$-matrix from the Hilbert space at
$\scri^-$ to that at $\scri^+$ using twistor methods (see, e.g.,
\cite{scattering1,scattering2}). Furthermore, using the BMS group one 
can assign mass and spin to these asymptotic quanta and show that they 
have $m=0$ and $s=2$. Thus, even in full general relativity the
elementary quanta can be identified with gravitons at null infinity.
Finally, one can show that there is a direct correspondence between
the `vacuum degeneracy' discussed in section \ref{s3.2} and infrared
issues in the quantum theory associated with `soft gravitons'.
However, these issues are much more complicated than in the Maxwell
case because of the second difficulty ---the `non-Abelian' character of general relativity \cite{aa-asym}--- and constitute a subject of ongoing investigations \cite{as,skinner,mcal}.

I will now provide a brief overview of all these quantum aspects. 
This discussion will include only those issues that are likely to
interest the geometric analysis community and will therefore not 
cover recent results, nor will the treatment of topics covered be 
comprehensive.

\subsection{Quantization of the radiative modes}
\label{s4.1}

The phase space $(\Gammabf, \Omegabf)$ of radiative modes provides a
natural point of departure for quantization. In field theories,
because of the presence of an infinite number of degrees of freedom,
von-Neumann's \cite{vn} celebrated uniqueness theorem (on
representations of the canonical commutation relations) is no longer
applicable. So we are led to an algebraic approach where one first
constructs the appropriate algebra of observables and then seeks its
representations on Hilbert spaces. In the case when the phase space
is linear ---as, e.g., in the Maxwell theory--- the `elementary'
observables generating these algebras are taken to be linear
functions on phase space. Their Hamiltonian vector fields are
constant on the phase space. Our phase space $\Gammabf$ is an affine
space and we can again seek functions whose Hamiltonian vector
fields are constant. It turns out that these can be obtained by
smearing the news tensor $N_{ab}$ by test fields. Let $f_{ab}$
denote symmetric second rank test fields on $\scri$, transverse to
$n^a$, that belong to the Schwarz space $\S$ of fields which, together 
with all their derivatives decrease rapidly (as one moves to $i^o$ and 
$i^\pm$ along $\scri^\pm$). Then the `elementary' classical observables 
on $\Gammabf$ turn out to be
\be \big[N[f]\big] (\{D\}) : = - \frac{1}{8\pi G}\, \int_\scri N_{ab}\,
f_{cd}\, q^{ac}q^{bd}\, \epsilon_{mnp}\, dS^{mnp}\, . \ee
The Hamiltonian vector field $X_{N[f]}$  is just the constant vector
field on $\Gammabf$, represented by $f_{ab}$. The Poisson brackets are given by
\be \{ N[f], \, N[f^{\prime}] \}_{\rm PB}\, = \, - \frac{1}{8\pi G}\,
\int_\scri [f_{ab}\, \Lie_{n}f^{\prime}_{cd} - f^{\prime}_{ab}\, \Lie_
{n}f_{cd}]\, q^{ac}q^{bd}\, \epsilon_{mnp}\, dS^{mnp}\,
 = \Omegabf(f,f^{\prime}) \, . \ee
Note that the right side is a constant. The algebra $\A$ of quantum
operators is generated by the (abstractly defined) $\hat{N}[f]$'s.
More precisely, $\A$ is the free $\star$-star algebra generated by
the $\hat{N}[f]$ subject to the relations
\be \hat{N}[f] +\lambda \hat{N}[f^{\prime}] = \hat{N}[f+\lambda 
 f^{\prime}]\, ; \quad \hat{N}^{\star}[f] = \hat{N}[f]; 
 \quad{\rm and}\quad [\hat{N}[f], \, \hat{N}[f^{\prime}]\,] = i\hbar\,\Omegabf(f,\,f^{\prime})\, \mathbb{I}\,  \ee
where $\lambda \in \mathbb{R}$ and $\mathbb{I}$ is the identity
operator. In finding representations, it is more convenient to work
with the Weyl Algebra $\W$, generated by $\hat{W}[f] = \exp [(i/\hbar)\, \hat{N}[f])$ because the vector space generated by $W[f]$ is closed 
under the product:
\be \hat{W}[f]\, \hat{W}[f^{\prime}] = e^{- \frac{i}{2\hbar}\, \Omegabf(f,\,f^{\prime)}}\,\,  \hat{W} [f+f^{\prime}]\, . \ee
To find its representations, one can use a standard construction due
to Gel'fand, Naimark and Segal (GNS) \cite{gns}: Given a positive
linear functional on $\W$, the GNS construction yields a canonical,
cyclic representation of $\W$ on a Hilbert space $\H$. The cyclic
state in $\H$ is called `the vacuum' (and denoted $|0\rangle$)
because any other state in the Hilbert space can be obtained by
acting repeatedly by elements of $\W$ on it. The positive linear
functional on $\W$ is called the \emph{vacuum expectation value
function} (or VEV) because in the GNS representation of $\W$ on
$\H$, it yields precisely the expectation values $\langle
0|\,\hat{W}[f]\,|0\rangle$ of $\hat{W}[f]$ in the state $|0\rangle$.

To select the VEV, it is simplest to introduce a K\"ahler structure
on the space $\S$ of test fields. By mimicking the procedure used
for Maxwell fields in Minkowski space-time, one is led to first
introduce a complex structure $J$ by decomposing the fields
$f_{ab}$ into positive and negative frequency parts, using the
affine parameter $u$ on the integral curves of $n^a$:\, $J\cdot
f_{ab} = i f^+_{ab} - i f^-_{ab}$ with
\be \label{J} f^+_{ab}(u,\theta,\phi) = \int_o^\infty \tilde{f}_{ab} (\omega,
\theta,\phi)\, e^{-i\omega u}\, \rmd \omega \quad {\rm and}\quad
f^-_{ab} = (f^+_{ab})^\star\, , \ee
where $u,\theta,\phi$ are the obvious coordinates on $\scri$. One can
readily verify that: i) $J$ does not depend on the particular
conformal frame chosen (i.e., $n^a$) or the coordinates used in its
construction; and, ii) it is compatible with the symplectic
structure $\Omegabf$. Thus, we have a K\"ahler space. Denote by $\h$
the Cauchy completion of $\S$ with respect to the resulting
Hermitian inner product
\be\label{ip} \langle\, f |\, f^{\prime}\rangle \, := \,
\frac{1}{2}\,\big(\Omegabf(f,\, Jf^{\prime}) \,+\, i\, 
\Omegabf (f,f^{\prime})\big)\, . \ee
This Hilbert space will feature prominently in what follows.

We can now define the required VEV on $\W$:
\be \langle 0|\,\hat{W}[f]\,|0\rangle\, :=\, e^{- \frac{1}{2\hbar}\,
\Omegabf(f,\, Jf)}\,  \ee
and carry out the GNS construction. The underlying representation
space $\H$ is naturally isomorphic to the symmetric Fock space based
on $\h$: \,
\be \H = \oplus_{n=0}^\infty\,\, \h_n^{\rm sym}  \ee
where for $n>0$, the Hilbert space $\h_n^{\rm sym}$ is the symmetric
tensor product of $n$ copies of $\h$ and $\h_0^{\rm sym} =
\mathbb{C}$. The normalized vector in $\mathbb{C}$ is the `vacuum', 
and elements of $\h_n^{\rm sym}$ are obtained by repeated action of
creation operators $a^\dag[f] = (1/2)(N[f] + i N[Jf])$ on the
vacuum. This representation of the Weyl algebra of news operators
provides the asymptotic Hilbert spaces at $\scri$ for the $S$-matrix
theory. Note that the underlying phase space $(\Gammabf, \Omegabf)$
played a key role both in the construction of the news operators and
in the introduction of the Ka\"hler structure that lies at the heart
of the representation.

\subsection{Properties of the representation and infrared issues}

Physically, it is convenient to tie this representation to the
classical phase space in order to make the correspondence between
the classical and quantum theories transparent. For definiteness,
let us fix a fiducial classical vacuum $\{ \mathring{D} \}_0$. One
can interpret the quantum vacuum state $|0\rangle$ as the coherent
state in $\H$, peaked at this configuration. Recall that any two
connections in the phase space $\Gammabf$ are related by a
symmetric, second rank, transverse tensor field $\sigma_{ab}$.
Consider the subspace $\Gammabf_{0}$ of $\Gammabf$ consisting of
connections $\{D\}$ which are related to the given $\{ \mathring{D}
\}_0$ by a $\sigma_{ab}$ which has \emph{finite norm} with respect
to (\ref{ip}). We can regard $\Gammabf_{0}$ as a vector space with
$\{ \mathring{D} \}_0$ as its origin and label every element $\{D\}$
of $\Gammabf_{0}$ with the corresponding tensor field $\sigma_{ab}$.
Then, there is a natural isomorphism, $\sigma_{ab} \to
f_{ab}=\sigma_{ab}$, between $\Gammabf_{0}$ and the 1-particle
Hilbert space $\h$. Recall that, in any Fock space, each 1-particle 
state defines a coherent state. The coherent state $\Psi_{f}$ is then 
the `canonical' semi-classical quantum state corresponding to the
classical phase space point $\{D\}$ labeled by $\sigma_{ab} = f_{ab}$. Note that because of the requirement of finiteness of norm,
$\Gammabf_{0}$ \emph{does not contain any classical vacuum}
$\{\mathring{D}\}$ other than $\{\mathring{D}\}_0$. Therefore, our
Hilbert space does not have any 1-particle states or coherent states
corresponding to the other classical vacua.

Recall from section \ref{s3.2} that each vacuum state selects a
Poincar\'e subgroup $\mathcal{P}$ of $\B$. It is easy to check that
the subgroup selected by $\{ \mathring{D} \}_0$ preserves the
1-particle Hilbert space $\h$. Furthermore, this action provides a
unitary, representation of that Poincar\'e group. One can show that
it can be decomposed into two irreducible representations. Every
irreducible representation of $\mathcal{P}$ is labeled by the
eigenvalues of its Casimir operators, mass and spin. An explicit
calculation shows that the eigenvalue of the mass operator is $0$ on
both irreducible representations and those of the spin operator are
$\pm 2$. When $m=0$, the spin vector is either parallel or
anti-parallel to the 4-momentum and the resulting configurations are
said to have positive and negative \emph{helicity} respectively. In
explicit terms, $f_{ab} \in \h$ has positive/negative helicity if its
positive frequency part is `self-dual/anti self-dual' in the sense that
$\epsilon^{mnp} \ell_p\, q_{nb} f^+_{am} = \mp i f^{+}_{ab}$ for any
$\ell_a$ on $\scri$ satisfying $\ell_an^a = -1$. (For details, see
\cite{aa-asym}, and for the simpler Maxwell example, \cite{aa-hel}).

The classical expressions of the fluxes $H_\xi$ of BMS momenta
(\ref{H}) can be promoted to self-adjoint quantum operators
$\hat{H}_\xi$ on $\H$. These operators provide us with the notion of 
fluxes of the BMS momenta in the quantum theory. It turns out that the
K\"ahler structure selected by the positive and negative frequency
decomposition guarantees that the answers one would get in the
quantum theory are fully compatible with those in the classical
theory. More precisely, if $|\,\Psi_{\{D\}}\,\rangle$ is the
coherent state in $\H$ peaked at a point $\{D\}$ in $\Gammabf_0$,
then
\be \langle\, \Psi_{\{D\}}|\,\hat{H}_{\xi}\, |\Psi_{\{D\}} \rangle =
H_{\xi} (\{D\})\, . \ee
Furthermore, this physical condition suffices to determine the
K\"ahler structure used in our quantization uniquely. These results
pave the way to establish that there exists a well-defined $S$-matrix 
theory on the sector of the asymptotic Hilbert space $\H$ that is 
spanned by coherent states peaked at the classical configurations $\{ D 
\} \in \Gammabf_0$ that are also sufficiently close to $\{ D\}_0$ in 
the sense that they are induced on $\scri^{\pm}$ by the evolution of 
`small initial data' \`a la \cite{dcsk,lb}.

If the data is not `small', the evolution may lead to the formation of 
a black hole in classical general relativity. In this case, $\scri^{+}$ 
would not serve as a good future boundary for the $S$-matrix theory. 
Furthermore, this can happen even when the incoming gravitons at $
\scri^{-}$ have completely `tame' frequencies. Therefore one would 
expect the $S$-matrix description to be inadequate in these 
circumstances, even though they occur well away from the Planck regime. 
This limitation was realized rather soon after the introduction of the 
asymptotic quantization program and drained the motivation from making 
a serious attempt to construct the quantum $S$-matrix at that time. 
Since then, advances in geometric analysis have furnished detailed 
results on the class of initial data that do lead to a well-defined $S
$-matrix on the classical side, and twistor methods are now enabling 
concrete progress on the quantum side. By combining these results, one 
may be able to obtain sharp results on the class of asymptotic states 
in $\H$ for which the quantum scattering operator is well defined to 
leading order in perturbation theory. This would resolve the 30 year 
old tension between black hole formation in the classical theory and 
the perturbative $S$-matrix program in the quantum theory.

Finally, the geometric structure at $\scri$ also sheds new light on
the infrared issues associated with gravity. Consider any point $\{
D \}$ in $\Gammabf_0$  labeled by the field $\sigma_{ab}$. One can
show that the news tensor $N_{ab}$ of $\{ D \}$ is given by $N_{ab}
= 2 \Lie_n \sigma_{ab}$. But recall that for $\sigma_{ab}$ to define
an element of $\h$ (i.e., a 1-graviton state), its norm under
(\ref{ip}) must be finite. This turns out to be a strong requirement,  for it implies
\be \label{ir} Q(\theta, \phi) := \int_{-\infty}^{\infty} \rmd u
\,N_{ab} (u, \theta, \phi)\, =0\, . \ee
If $Q(\theta,\phi) \not=0$, the norm (\ref{ip}) is infrared divergent 
(i.e. diverges because the integral in (\ref{J}) extends to $\omega 
=0$). Hence $Q(\theta, \phi)$ is referred to as the `infrared charge' 
of $\{D\}$. Note that mathematically (\ref{ir}) amounts to an infinite
number of conditions because the integral over $u$ has to vanish for 
each generator of $\scri$, labeled by $\theta,\phi$. To understand its 
physical content, it is best to recast it in terms of connections. It 
is straightforward to show that (\ref{ir}) is equivalent to asking 
that, for a connection $\{ D\}$ to be in $\Gammabf_0$, it must approach 
\emph{the same} vacuum configuration at both ends of $\scri$. (On $
\scri^+$ the two ends are $i^o$ and $i^+$ and on $\scri^-$ they are 
$i^-$ and $i^o$.) As we discussed in section \ref{s3.2}, because of the 
memory effect \cite{memory2}, this is physically too stringent a requirement. In 
scattering processes one does not expect it to be satisfied except in 
rather exceptional circumstances.

This means that, physically, it would be unreasonable to restrict
oneself to the Fock representation we have constructed above. However, 
it is relatively straightforward to enlarge the asymptotic Hilbert 
spaces by allowing suitable `displaced Fock representations' 
\cite{aa-asym}. This enlargement is necessary already in quantum 
electrodynamics where, even after renormalization, we do not have a 
well-defined $S$-matrix order by order in perturbation theory because 
of the infrared problems associated with photons which are completely 
analogous when formulated in terms of $\scri$. In that case, one can find the `minimal enlargement' to make the $S$-matrix well-defined in an efficient and elegant manner \cite{aa-asym,aakn}. However, the procedure makes crucial use of the fact that the photons themselves do not act as sources of the quantum Maxwell field; the theory is an Abelian gauge theory. Although this simplification is not available in gravity, there have been interesting recent advances within perturbation theory \cite{jwrsrk}. It would be interesting to investigate if  these results can lead to a coherent quantum theory of the `small data' sector for which we have global existence theorems in the classical theory.

\section{Discussion}
\label{s5}

Structure of null infinity is both subtle and rich. In the classical theory, it has led to some surprising results. In particular, there is a `vacuum degeneracy' in the phase space of connections directly related to the presence of supertranslations, i.e., to the fact the asymptotic symmetry group is the infinite dimensional BMS group $\B$ rather than the 10-dimensional Poincar\'e group. The BMS group acts via symplectomorphisms on the phase space and their  Hamiltonians provide us with the expressions of the BMS momenta across $\scri$ that are important for the physics of gravitational waves. This review focused on the geometric aspects of null infinity, rather than on global issues on the existence and uniqueness of solutions to Einstein's equations. A summary of the rich set of results in that area can be found, e.g., in \cite{dcsk,hf1,lb,hf2,sair}.

In the quantum regime, the structure at $\scri$ enables one to construct suitable operator algebras and their representations. The Hilbert space underlying these representations  provides the asymptotic states for the $S$-matrix theory. This framework provides a precise sense in which gravitons arise in full non-linear gravity, without having to make expansions around Minkowski space-time. Furthermore, since one can introduce abstract boundaries $\scri^{\pm}$ equipped with structures that are sufficiently rich to carry out these constructions without reference to the space-time interior, one can now envisage various space-time geometries as paths, or histories, that interpolate between the fixed, well-defined asymptotic states on $\scri^{\pm}$. Therefore, there is also a potential for calculating a \emph{non-perturbative} sum over histories, e.g., along the lines of spin foams (see, e.g., \cite{arr}), or, by allowing classical geometries with different topologies in the bulk in the spirit of Wheeler's `space-time foam' paradigm.

In both regimes interesting open issues still remain at the interface of geometric analysis and gravitational physics. Perhaps the most surprising among these  is a set related to the cosmological constant $\Lambda$. Every issue discussed in this overview assumes $\Lambda=0$, while by now we have very strong indications from observations that $\Lambda$ is small but positive. At first, one might expect that it would be trivial to extend the interplay between geometry and physics to accommodate the presence of a non-zero $\Lambda$. And indeed, there have been a number of interesting results for the $\Lambda<0$ case. For $\Lambda >0$, a global existence result, due to Friedrich \cite{hf3}, has been available for a long time. However, from a physical perspective, very little is known in the $\Lambda >0$ case. As of now, we do not have a physically useful positive energy theorem because, for $\Lambda >0$, the asymptotic symmetry representing  `time translation' that defines energy is space-like near $\scri$. We also do not have the analog of the Bondi news. So, \emph{we do not even have a gauge invariant characterization of gravitational waves in full general relativity,} let alone the phase space of radiative modes and formulas for Hamiltonians representing fluxes of energy-momentum! Thus, even a tiny cosmological cosmological constant casts a long shadow, if it is positive \cite{aabbak}, as in the real universe. Extensions of the rich results outlined in this overview to the $\Lambda >0$ case represent a vast field of opportunities for new, interesting results for both physics and mathematics communities. I hope that this overview will help enhance the interaction between the two.

\section*{Acknowledgments}
This work was supported in part by the NSF grants PHY-1205388,  PHY-1505411 and the Eberly research funds of Penn State. I thank B\'{e}atrice Bonga and Aruna Kesavan for a careful reading of the manuscript and the referee for useful suggestions.

\end{document}